\documentclass[aps,pra,twocolumn,showpacs,floatfix,longbibliography]{revtex4-2}
\usepackage{graphicx,amsfonts,amssymb,amsmath}
\usepackage{textcomp}

\usepackage{footnote}
\usepackage{esint}
\usepackage[english]{babel}
\usepackage{afterpage}
\usepackage{bbold}
\usepackage{soul}
\usepackage[colorlinks,linktocpage,citecolor=blue,linkcolor=red,urlcolor=blue]{hyperref}

\input cyracc.def

\allowdisplaybreaks

\newtheorem{remark}{Remark}

\begin{document}
	
	
	
\title{Time-translation invariance symmetry breaking hidden by finite-scale singularities}

\author{Ixandra Achitouv}
\email[]{ixandra.achitouv@cnrs.fr}
\affiliation{Institut des Syst\`emes Complexes ISC-PIF , CNRS, 113 rue Nationale,
Paris, 75013, France.}

\author{Vincent Lahoche}
\email[]{vincent.lahoche@cea.fr}
\affiliation{Université Paris Saclay, \textsc{Cea}, Gif-sur-Yvette, F-91191, France}

\author{Dine Ousmane Samary}
\email[]{dine.ousmanesamary@cea.fr}
\affiliation{Université Paris Saclay, \textsc{Cea}, Gif-sur-Yvette, F-91191, France}
\affiliation{Faculté des Sciences et Techniques (ICMPA-UNESCO Chair)\protect\\
Université d'Abomey-Calavi, 072 BP 50, Bénin}

\author{Parham Radpay}
\email[]{parham.radpay@cea.fr}
\affiliation{Université Paris Saclay, \textsc{Cea}, Gif-sur-Yvette, F-91191, France}

\begin{abstract}
\medskip

In this paper, we consider a renormalization group perspective on the quantum dynamics of a particle moving in the Euclidean $\mathbb{R}^N$ space through the complex landscape provided by a disordered Hamiltonian of type ‘‘$2+p$". We focus on the large $N$ limit, where the coarse-graining procedure is unconventional: it is based on the Wigner spectrum of the rank-2 disorder.
The main consequence of this choice is that canonical dimensions depend on the scale, and the flow equations fail to become autonomous, preventing the existence of global fixed points. One of the main features of the underlying renormalization group flow is the existence of finite-scale singularities for initial conditions sufficiently close to the Gaussian region and for rank-$p$ disorder intensity large enough. Using the Luttinger-Ward formalism, we show that these finite-scale singularities hide (and should be resolved by) a phase transition that breaks time-translation invariance.
\medskip

\end{abstract}
	
\maketitle
 

\section{Introduction} 
Equilibrium and out-of-equilibrium glassy systems are essentially characterized by the complexity of their energy landscape. Classically, and below the critical temperature, the system is trapped in a very large number of metastable states for a long time \cite{Dotsenko,Castellani,Mezard,Dominicis}. Usually, the transition toward a glassy regime can be characterized using different signatures, such as replica symmetry breaking, weak ergodicity breaking, or complexity singularity. However, there is no indication as to which one of these approaches is the best.

More concerning is that the computational frameworks underlying these indicators predict different critical temperatures. For this reason, they can be viewed as incomplete methods, capturing some aspects but not all of the physics of the transition \cite{Castellani}. Moreover, it is important to note that despite their quantitative disagreements, these methods remain debated mathematically, especially regarding the construction of the $n \to 0$ limit, possibly leading to the replica replica symmetry breaking \cite{Agliari,Mezard,Dotsenko}. \medskip

For historical and analytical  reasons, the renormalization group (RG) is not one of the main topics for studying glassy systems; its relevance accounts for only a small percentage of publications about spin glass and glassy physics. The reason underlying this phenomenon likely stems from the origins of RG, which emerged from critical phenomena and was motivated primarily by the breakdown of mean-field theory for weakly connected spins in low dimensions. In contrast, spin glass systems are generally highly connected, for which mean-field theory works well. However, the fact that mean-field theory remains a good approximation does not disqualify the RG. The question is not whether one method works better than another but whether we can derive reliable information from it. In this respect, RG can—and in fact already has—provided a significant framework for studying glassy systems \cite{Pimentel,Castellana,Bray,Parisi,Biroli,Tarjus1,Tarjus2,Tarjus3,Tarjus4,Dominicis} (the list is far from exhaustive). This paper follows a series \cite{Lahoche1,Lahoche2,Lahoche3}, and our main goal is to develop reliable approximations for models that are easy to benchmark with standard tools, with the aim of investigating more complex systems where other methods become difficult to track. \medskip

In this paper, we consider a specific class of glassy systems as a suitable formalism to investigate quantum effects in spin glasses. In general spin glass systems, the transition temperatures $T_c$ are large enough that the energy scale $k_B T_c$ obscures quantum effects. However, this may not be the case for certain values of experimentally controlled extra-parameters, such as pressure or magnetic fields, on which the critical temperature depends. As soon as the critical temperature becomes small enough, quantum effects are expected to play a significant role in understanding the transition—this is the topic of quantum spin glasses. One of the main relevant effects that could explain a quantitative difference between the quantum and classical cases is the influence of tunneling effects. \medskip

In the literature, there are different mathematical approaches to understanding the quantum nature of glassy systems. The most natural one is the disordered version of the so-called quantum Heisenberg model, where Ising spins in the Hamiltonian are quantized using Pauli matrices \cite{Cugliandolo3}. Other aspects of quantum properties can be addressed from different perspectives. For instance, the standard classical spherical $p$-spin model can be viewed as a classical particle moving on an $N$-dimensional hypersphere under the influence of a random potential. Quantization can be constructed by imposing Heisenberg commutation relations between conjugate classical variables or, as we propose in this paper, using Feynman's path integral formalism \cite{Cugliandolo1,Cugliandolo2}. More precisely, the model we consider is not spherically constrained; instead, the hard spherical constraint usually imposed by a Dirac delta function is replaced by a “gentle" polynomial potential that avoids classical large-spin configurations. \medskip

In a series of papers \cite{Lahoche1,Lahoche2,Lahoche3}, we investigated this quantum version of the classical $p$-spin models through the lens of the functional renormalization group (FRG), focusing on the Wetterich formalism \cite{Wetterich1,Wetterich2,Wetterich3}. We considered two types of spin glasses and adopted two different strategies. On one hand, the $2+p$ quantum spin glass involves two types of disorder: an $N \times N$ matrix-like disorder represented by a random matrix and a rank-$p$ random tensor.

This model was studied classically, and its fully replica symmetry-breaking solution was constructed in \cite{Leuzzi}. In \cite{Lahoche1,Lahoche2}, focusing on the large $N$ limit, we constructed an RG approach based on coarse-graining over the Wigner spectrum, with four key observations: \begin{enumerate} \item The canonical dimensions for couplings depend on the scale, but the theory behaves as a standard 4D non-local field theory in the deep infrared (IR). \item Finite-scale singularities appear in the RG flow near the Gaussian fixed point for sufficiently large rank-$p$ disorder. \item A second non-vanishing minimum appears for the Luttinger-Ward functional associated with the $2$-point correlation $q_{\alpha\beta}$ between replicas for divergent trajectories, suggesting a first-order phase transition. \item Divergences are canceled in some regions of the full phase space as truncations are “enhanced" by local interactions coupling different replicas. \end{enumerate} The last point was the main focus of our previous work \cite{Lahoche2}. On the other hand, we considered the $p$-spin quantum spin glass from a more conservative perspective. In \cite{Lahoche3}, we constructed a coarse-graining over time (frequencies), which differs from the Wigner spectrum case and corresponds to an outer scale of the field theory. The resulting formalism is that of the RG for a standard field theory in dimension 1, where interactions can be local or non-local. Field theories in dimension 1 are inherently pathological since all interactions are relevant, making the construction of approximate solutions more difficult. Nevertheless, we obtained similar results through different approximations, which lends a certain robustness to these findings. Quite surprisingly, these results agreed with those obtained by coarse-graining over the Wigner spectrum in \cite{Lahoche2}. In particular, we again observed the presence of a singularity at a finite (time) scale, which seems to induce a state of phase coexistence between a state with no correlation between replicas and a state involving correlations (but without magnetization). Finally, incorporating operators that couple the replicas removes the singularities in certain regions of the phase space. \medskip

This mechanism resembles observations in other contexts (see, for example, \cite{Tarjus1,Tarjus2,Delamotte,Dupuis}) and could have been expected. However, the fact that both approaches (time v.s. Wigner coarse-graining) yield similar results may seem mysterious and invites deeper reflection on the role of time. Another question concerns the possibility that other correlations, absent in perturbation theory, may play a role in other regions of the phase space. In our previous work, we primarily considered local correlations between replicas, but this does not completely define the phase space, as suggested by the persistence of some divergences. In this note, we investigate the possibility of phase transitions to regimes where time-translation invariance is broken, with or without correlations between replicas. In this paper, we focus on a ‘‘2+p" type model, building a coarse-graining on the Wigner spectrum, reserving a more detailed global analysis for future work. Time translation invariance characterizing
 equilibrium dynamics generally does not
hold for glassy systems, signaling
that the system is not able to equilibrate even at long times. These effects occur also in classical spin glasses, and are in particular responsible of aging effects \cite{Cugliandolo1,Cugliandolo2,Dominicis}.
\medskip

\noindent \textbf{Outline.} The outline of the paper is as follows. In Section \ref{sec1}, we introduce the model, conventions, and basic definitions. In Section \ref{sec2}, we consider the 1PI formalism (non-perturbative RG) and the canonical scaling along the Wigner spectrum, and investigate approximate solutions. In Section \ref{sec3}, we examine the Luttinger-Ward (LW) functional for the $2$-point function, including operators that explicitly break time-translation symmetry. By investigating the behavior of the LW functional just above and below the critical value for the rank-$p$ disorder, we show that finite-scale singularities “hide" a first-order phase transition toward a phase without time-translation symmetry. \medskip

\noindent \begin{remark} For the reader, let us note that although we focus primarily on the case $p=3$ in this paper, as in \cite{Lahoche1}, the formalism generalizes trivially to the disorder of larger rank. The price to pay is the additional complexity of the flow equations. Our primary goal in this first series of articles is to demonstrate a concept, so we limit ourselves to the first non-trivial case. \end{remark}

\section{The model}\label{sec1}

The model we focus on is a slightly modified version of the quantum spin glass model considered in \cite{Cugliandolo1,Cugliandolo2}. We consider a quantum particle moving in Euclidean space $\mathbb{R}^N$. The wave function $\psi(\textbf{x}, t)$, for $\textbf{x} \in \mathbb{R}^N$, satisfies the Schrödinger equation:
\begin{equation}
\boxed{-\hbar\frac{d}{dt}\, \psi(\textbf{x},t)=\hat{\mathcal{H}}(\textbf{x},t) \psi(\textbf{x},t)\,,}
\end{equation}
where the quantum Hamiltonian $\hat{\mathcal{H}}(\textbf{x},t)$ is:
\begin{equation}
\hat{\mathcal{H}}(\textbf{x},t):=-\frac{\hbar^2}{2 m_0} \frac{\partial^2}{\partial \textbf{x}^2} + U(\textbf{x}^2) + V_{J,K}(\textbf{x})\,,
\end{equation}
where $t\in [-\beta/2,\beta/2]$, $U$ is some polynomial function with argument the single $O(N)$ invariant $\textbf{x}^2:=\sum_{i=1}^N x_i^2$, and:
\begin{equation}
V_{J,K}(\textbf{x}):=\frac{1}{2} \sum_{i,j} K_{ij} x_i x_j+\sum_{i_1\leq \cdots \leq i_p} J_{i_1\cdots i_p}x_{i_1}\cdots x_{i_p}\,.
\end{equation}
In this equation $K$ and $J$ are quenched random couplings, $K$ is a Wigner matrix \cite{RMT} of size $N$ and variance $\sigma/N$, and $J$ is a Gaussian random tensor with zero mean and variance:
\begin{equation}
\overline{J_{i_1\cdots i_p}J_{i_1^\prime \cdots i_p^\prime}}=\left(\frac{\kappa^2 p!}{N^{p-1}}\right) \,\prod_{\ell=1}^p \delta_{i_\ell i_\ell^\prime}\,.\label{averageJ}
\end{equation}
We use the notation $\overline{X}$ for the average over rank $p$ disorder distribution. The most celebrated theorem of random matrix theory states that eigenvalues $\lambda_\mu \in \mathbb{R}$ of the Wigner matrix $K$ in the large $N$ limit are distributed accordingly with the so-called \textit{Wigner semicircle}:
\begin{equation}
\mu_W(\lambda):=\frac{\sqrt{4\sigma^2-\lambda^2}}{2\pi \sigma^2}\,.\label{Wigner}
\end{equation}
This result, for the computation we have done in this paper, means that for gentle enough function $f(x)$, discrete sums can be replaced by integrals:
\begin{equation}
\frac{1}{N}\sum_{\mu=1}^N \, f(\lambda_\mu) \to \int_{-2\sigma}^{+2\sigma} d\lambda \mu_W(\lambda) f(\lambda). 
\end{equation}
The positive real number $\kappa^2$ measures the magnitude of the disorder and is expected to be smaller than the variance $\sigma^2$ of matrix-like disorder. In this limit, the system can be expected to behave almost like the quantum $p=2$ model, which is exactly solvable \cite{Dominicis}. The solution resembles what is obtained for the classical problem: a second-order phase transition occurs at sufficiently low temperatures, where the component $\mu = -2\sigma$ of the projection $x_\mu$ along the eigenvector corresponding to the eigenvalue $\lambda_\mu$ has a macroscopic occupation number, $(x_{-2\sigma})^2 \sim \mathcal{O}(N)$.

If $\sigma^2 \gg \kappa^2$, the effects of the disorder are expected to superpose onto a background field that essentially corresponds to the low-temperature $2$-spin glass phase. The coarse-graining we construct implicitly assumes this limit and provides a first approximation of the RG flow around the corresponding low-temperature vacuum, as described in our previous work \cite{Lahoche1}.

This work focuses exclusively on the high-temperature symmetric phase, assuming that the potential's concavity remains unchanged.
\medskip

The quantization can be done using standard path integral approaches for a given realization of the disorder's $K$ and $J$. Physically, it corresponds to the partition function of a quantum particle in contact with a thermal bath at temperature $\beta^{-1}$:
\begin{equation}
\mathcal{Z}_{\beta}[K,J,\textbf{L}]=\int [\mathcal{D} x(t)]\, e^{-\frac{1}{\hbar} S_{\text{cl}}[\textbf{x}(t)]+\frac{1}{\hbar} \int dt \sum_{k=1}^N L_k(t) x_k(t)}\,,\label{pathintegralZ}
\end{equation}
where $[\mathcal{D} x(t)]$ denotes the path integral ``measure'', $\textbf{L}=(L_1,\cdots, L_N)$ is the source field and where the \textit{classical action} $S_{\text{cl}}[\textbf{x}(t)]$ is:
\begin{equation}
S_{\text{cl}}[\textbf{x}(t)]:=\int_{-\beta/2}^{\beta/2} dt \left(\frac{1}{2}\dot{\textbf{x}}^2+V_{J,K}({\textbf{x}})+U({\textbf{x}}^2)\right)\,,\label{classicaction}
\end{equation}
provided with periodic boundary conditions $\textbf{x}(t)=\textbf{x}(t+\beta)$. In this paper, we essentially focus on the limit, $\beta\to \infty$ i.e. for vanishing temperature. Note that in the large $N$ limit and the symmetric phase the physical mass does not renormalize, and we set $m_0=1$. As is usual in quantum field theory, the partition function allows computing time ordered vacuum-vacuum expectation value of field correlations at different times \cite{Weinberg} $\langle 0 \vert T\,\hat{x}_{i_1}(t_1)\hat{x}_{i_2}(t_2)\cdots \hat{x}_{i_n}(t_n) \vert 0 \rangle$, where the hat means quantum operator.
\medskip

Let us discuss the averaging process before the Functional Renormalization Group (FRG) analysis. The construction of the disorder averaging in the quenched regime is a technically challenging and still debated topic in physics and mathematics. The most popular method remains the replica method, which involves averaging over the disorder for $n$ copies of the original system. Typically, this method assumes an analytic continuation to construct the limit $n \to 0$, which allows for the possibility of replica symmetry breaking.

In the FRG literature, however, a different approach is often preferred, where the replica symmetry is explicitly broken from the outset by choosing different sources $\textbf{L}_{\alpha}$, with $0 \leq \alpha \leq n$, for each replica. Moreover, in the large $N$ limit, the random spectrum of the matrix $K$ converges to the deterministic Wigner semicircle. This convergence makes it advantageous to exploit the underlying $O(N)$ invariance of the problem (the probability measures for $J$ and $K$ are expected to be $O(N)$-invariant) and work in the eigenbasis of $K$, thereby avoiding the need to average over the matrix-like disorder. Schematically \cite{Lahoche2}:
\begin{equation}
\mathcal{Z}_{\beta}[K,J,\textbf{L}] \to \tilde{\mathcal{Z}}_{\beta}[\mu_W,J,\textbf{L}]\,,
\end{equation}
where, on the right-hand side, we assume that discrete sums are replaced everywhere by integrals in the computation of Feynman diagrams for perturbation theory. Finally, we will consider the averaging:
\begin{equation}
\boxed{\bar{Z}_\beta[\mu_W,\{\textbf{L}_\alpha\}]:= \overline{\prod_{\alpha=1}^n \tilde{\mathcal{Z}}_{\beta}[\mu_W,J,\textbf{L}_\alpha]}\,.}\label{averagedZ}
\end{equation}
This is this averaged functional, for an arbitrary value of $n$, which is generally considered in the FRG literature \cite{Castellani}, aiming to focus on the cumulant of the random large $N$ free energy $\ln \tilde{\mathcal{Z}}_{\beta}[\mu_W,J,\textbf{L}]$. The classical action for the averaged replicated theory is:
\begin{align}
\nonumber\overline{S_{\text{cl}}}&[\{\textbf{x}_\alpha\}]:=\sum_{\alpha}S_{\text{cl}}[\textbf{x}_\alpha(t),J=0,K]\\
&\quad -\frac{\kappa^2 N}{2\hbar}\int_{-\beta/2}^{+\beta/2} d t \, d t^\prime\sum_{\alpha,\beta}\,  \left(\frac{\textbf{x}_\alpha(t)\cdot \textbf{x}_\beta(t^\prime)}{N}\right)^p\,.\label{classicalaveraged}
\end{align}
In the rest of this paper, we set $\hbar=1$ everywhere. Moreover, we consider only the case $p=3$ for explicit calculations.

\section{1PI formalism and finite scale singularities}\label{sec2}

\subsection{Wetterich formalism}
This section introduces the one-particle irreducible (1PI) Wetterich formalism for the averaged theory we considered just above \eqref{averagedZ}. In the Gauge where the matrix-like disorder is diagonal, the kinetic kernel is, in the Fourier space:
\begin{equation}
\mathcal{K}=\omega^2+\lambda_\mu+2 U'(0)=: \omega^2+p_\mu^2+m^2\,,
\end{equation}
where $p_\mu^2:=\lambda_\mu+2\sigma$, $m^2:=2 U'(0)-2\sigma$. In the large $N$ limit, $p_\mu^2$ becomes positive definite and looks as a \textit{generalized momentum}. This is along the spectrum of this momentum that we will perform the coarse-graining. The standard strategy \cite{Wetterich1} is to add some regulator $\Delta S_k$ to the classical action $S_{\text{cl}}$ \eqref{classicaction}, suppressing infrared degrees of freedom from long-range physics. In this paper, we consider the following: 
\begin{equation}
\Delta S_k[\{\textbf{x}_\alpha\}]:= \frac{1}{2}\sum_{\alpha=1}^n \sum_{\mu=1}^N \int dt\, x_{\alpha \mu}(t) R_k(p_\mu^2)x_{\alpha i}(t)\,.
\end{equation}
In this paper, we will focus on the slightly modified Litim regulator, $R_k(p^2):=f(k)(k^2-p^2)\theta(k^2-p^2)$. The regulator is designed such that $k\in [0,4\sigma]$ interpolates between the classical action $\overline{S_{\text{cl}}}$ in the deep ultraviolet ($k\to 4 \sigma$), and the effective action $\Gamma$ i.e. the Legendre transform of the replicated free energy $\ln \bar{Z}_\beta$ in the deep infrared ($k\to 0$). For the function $f(k)$ we choose as in \cite{Lahoche1}:
\begin{equation}
f(k):=\frac{4\sigma}{4\sigma-k^2}\,.
\end{equation}
The interpolation can be converted as a differential equation describing the trajectory of the system on the action space, and in the 1PI formalism is the \textit{Wetterich equation} \cite{Wetterich1}:
\begin{equation}
\boxed{\dot{\Gamma}_k=\frac{1}{2} \mathrm{Tr} \big\{ \dot{R}_k \big(\Gamma_k^{(2)}+R_k\big)^{-1} \big\}\,.}\label{Wett}
\end{equation}
where $\Gamma_k^{(2)}$ is the second-order functional derivative of $\Gamma_k$, the dot means derivative with respect to $t:=\ln (k/4\sigma)$ and the trace involves a sum over momenta and frequencies as well as the replica indices. The precise definition of the functional $\Gamma_k$ is the following:
\begin{align}
\nonumber \Gamma_k[\{\textbf{M}_\alpha \}]+\ln \bar{Z}_\beta[\mu_W,\{\textbf{L}_\alpha\}]&=\sum_{\alpha=1}^n\, \int dt \, \textbf{L}_\alpha(t) \cdot \textbf{M}_\alpha(t)\\
&-\Delta S_k[\{\textbf{M}_\alpha\}]\,,
\end{align}
the \textit{classical field} $\textbf{M}_\alpha$ is defined as:
\begin{equation}
\textbf{M}_\alpha(t):=\frac{\delta}{\delta \mathbf{L}_\alpha(t)} \ln \bar{Z}_\beta[\mu_W,\{\textbf{L}_\alpha\}]\,.
\end{equation}
The flow equation \eqref{Wett} operates in the infinite-dimensional theory space. While it is formally exact, solving it exactly is impossible. To extract nonperturbative information about the flow, approximations—commonly referred to as truncations—are employed. These take the form of a suitable ansatz for $\Gamma_k$.

Since we focus on the symmetric phase, $\Gamma_k$ should be expanded as a power series in the field. One way to circumvent the technical difficulty of inverting the $n \times n$ matrix $\Gamma_k^{(2)}$, while aligning with our proposal to focus on the cumulants of the random distribution $\ln \tilde{\mathcal{Z}}{\beta}[\mu_W, J, \textbf{L}]$, is to expand $\Gamma_k[{\textbf{M}\alpha }]$ in terms of an increasing number of free replica sums and local components \cite{Dupuis, Tarjus1, Lahoche4}.
\begin{align}
\nonumber &\Gamma_k[\{\textbf{M}_\alpha \}]=\int dt \sum_{\alpha}\gamma_{k,1,1}[\{\textbf{M}_\alpha(t)\}]\\\nonumber
&+\frac{1}{2}\int dt dt^\prime \sum_{\alpha} \gamma_{k,2,1}[\{\textbf{M}_\alpha(t)\},\{\textbf{M}_\alpha(t^\prime)\}]\\\nonumber
&+\frac{1}{2}\int dt \sum_{\alpha,\beta} \gamma_{k,1,2}[\{\textbf{M}_\alpha(t)\},\{\textbf{M}_\beta(t)\}]\\\nonumber
&+\frac{1}{4}\int dt dt^\prime \sum_{\alpha,\beta} \gamma_{k,2,2}[\{\textbf{M}_\alpha(t)\},\{\textbf{M}_\beta(t^\prime)\}]\\\nonumber
&+\frac{1}{3!}\int dt dt^\prime dt^{\prime\prime} \sum_{\alpha}\gamma_{k,3,1}[\{\textbf{M}_\alpha(t)\},\{\textbf{M}_\alpha(t^\prime)\},\{\textbf{M}_\alpha(t^{\prime\prime})\}]\\
&\,+\,\cdots\,.
\end{align}
Each multi-local component $\gamma_{k,m,n}$ involves $m$ different replicas and $n$ local components. We denote the corresponding component in the expansion of $\Gamma_k$ as $\Gamma_{k,m,n}$. For standard local field theory, the expansion terminates at $\gamma_{k,1,1}$, which is expanded as a power series in the classical field in the symmetric phase. In this case, the expansion takes the form of a sum of products of fields at the same time, and the various terms in the flow equations, expanded in powers of the classical field, can be indexed by effective Feynman graphs.

\begin{figure}
\begin{center}
\includegraphics[scale=1.2]{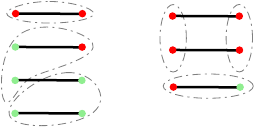}
\end{center}
\caption{A typical octic interaction contributing to $\Gamma_{k,3,2}$ (on left) and a typical sextic contribution (on right).}\label{hypergraph}
\end{figure}

In contrast, to account for non-localities and free replica sums, the flow equations are now indexed by hypergraphs \cite{Tegge} instead of ordinary graphs. A typical monomial in the expansion of $\Gamma_{k,m,n}$ involves $2P$ fields, $m$ local clusters where the fields interact at the same time, and $n$ clusters sharing the same replica index. Such a vertex can be represented as shown in Figure \ref{hypergraph}, which illustrates the rule for an octic interaction contributing to $\Gamma_{k,3,2}$.
The rule is as follows:
1) Fields are represented by nodes, connected pairwise by solid heavy lines that represent scalar products (contraction of Latin indices).
2) Fields sharing the same replica index are represented by nodes of the same color.
3) Fields interacting at the same time are enclosed within a dash-dotted closed path.
Thus, the octic vertex $v_{3,2}$, depicted in Figure \ref{hypergraph}, is explicitly written as:
\begin{align}
\nonumber v_{3,2}&:=\int \prod_{\ell=1}^3 dt_\ell \sum_{\alpha,\beta}\, \textbf{M}_{\alpha}(t_1)\cdot  \textbf{M}_{\alpha}(t_1)\times \textbf{M}_{\alpha}(t_2)\cdot  \textbf{M}_{\beta}(t_2)\\
&\times \textbf{M}_{\beta}(t_2)\cdot  \textbf{M}_{\beta}(t_3) \times \textbf{M}_{\beta}(t_3)\cdot  \textbf{M}_{\beta}(t_3)\,,
\end{align}
the dot corresponding to the Euclidean scalar product $\textbf{x}\cdot \textbf{y}:=\sum_{i=1}^N x_i y_i$. 

This approximation generalizes the standard local potential approximation, corresponding to the component $\Gamma_{k,1,n}$. Here, we allow the existence of multiple local clusters, but we do not take into account derivative interactions within these clusters. The flow equations for the different couplings are then obtained from the exact flow equation \eqref{Wett} and the ansatz for $\Gamma_k$, by identifying on the right and left-hand sides the components having the same number of local components, free replica sums and field involved on each of these components. 

\subsection{Canonical dimensions and flow equations}
This section takes up in the context of this article many results derived in \cite{Lahoche1,Lahoche2}, to which the reader can refer.

\begin{figure}
\begin{center}
\includegraphics[scale=1.1]{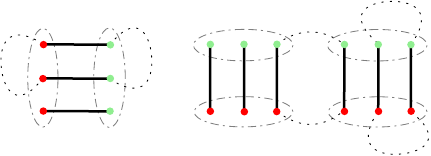}
\end{center}
\caption{Hypergraphs contributing to the non-local $2$-points function renormalization (on left) and on the quartic disorder renormalization (on right). Both do not create enough closed paths (faces) and are next to leading ($\mathcal{O}(1/N)$).}\label{fig2}
\end{figure}

The bare theory reads ($p=3$):
\begin{align}
\nonumber &\Gamma_{k=2\sqrt{\sigma}}[\{\textbf{M}_\alpha \}]=\int dt \sum_{\mu,\alpha} \, M_{\alpha \mu}(t) \left(-\frac{d^2}{dt^2}+p_\mu^2\right)M_{\alpha \mu}(t)\\\nonumber 
& +\, \vcenter{\hbox{\includegraphics[scale=0.85]{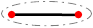}}}
\,+\,\vcenter{\hbox{\includegraphics[scale=0.8]{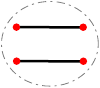}}}\,+\,\vcenter{\hbox{\includegraphics[scale=0.75]{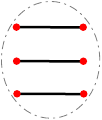}}}\,+\,\cdots +\,\vcenter{\hbox{\includegraphics[scale=0.8]{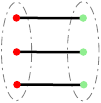}}}\,.
\end{align}
The bare propagator is local in time (i.e. proportional to $\delta(t-t^\prime)$) and diagonal in the replica space. Moreover, the existence of the large $N$ limit imposes that interactions involving $2P$ fields must scale as $N^{-P+1}$. Hence, in the large $N$ limit, the theory space connected to the Gaussian region (i.e. where the perturbation theory holds) from the RG flow is highly constrained. In particular, we find that the non-local interaction coming from the averaging over disorder does not renormalizes and that the self-energy remains a local function, diagonal in the replica space, the non-local correction being next to leading order comparing to local contributions (see Figure \ref{fig2}). For the same reason, no anomalous dimension is expected at the leading order \cite{Lahoche1}. 

Hence, assuming to work in the symmetric phase, perturbation theory invites to consider the following truncation:
\begin{align}
\nonumber \Gamma_k[\{\textbf{M}_\alpha \}]&=\frac{1}{2}\int dt \sum_{\mu,\alpha} M_{\mu,\alpha}(t)(-\partial_t^2+p_\mu^2+\mu_1)M_{\mu,\alpha}(t)\\\nonumber
&+\sum_{n=2}^\infty\int dt \sum_{\mu,\alpha} \frac{(2\pi)^{n-1}u_{2n}}{(2n)!N^{n-1}}\, \bigg(\sum_\mu M_{\mu,\alpha}^2(t) \bigg)^n\\
&+\frac{(2\pi)\tilde{u}_6}{6!N^2}\int dt dt^\prime \sum_{\alpha,\beta}\, \bigg(\sum_\mu M_{\mu,\alpha}(t) M_{\mu,\beta}(t^\prime) \bigg)^3\,,\label{truncationGamma}
\end{align}
which also defines coupling constants. The flow equations for the coupling constants involve effective loops, integrating over both the frequencies (in the limit $\beta \to \infty$) and the generalized momentum $p^2$ distributed according to Wigner's law. In an ordinary field theory, this type of integral behaves as a power of $k$, which can be eliminated in favor of an appropriate scaling of the couplings which then become dimensionless. Here, the integral depends non-trivially on $k$, and a simple power-law scaling is no longer sufficient to eliminate the dependence on the IR cut-off $k$ of the loop integrals. An additional complication comes from the fact that this integral also depends non-trivially on the mass. In this paper, we will focus on a regime where the mass is small enough to be neglected (which we will verify in our simulations). Under these conditions, each integral receives the contribution (see \cite{Lahoche1} for more details)
\begin{equation} 
\Omega(k):=k^3\int dp^2\, \frac{\rho(p^2) \dot{R}_k(p^2)}{(p^2+R_k(p^2))^{3/2}}=\frac{8 k^5}{3 \pi  \sqrt{4-k^2}}\,,
\end{equation}
and the \textit{dimensionless} local couplings are defined as:
\begin{equation}
\bar{u}_{2n}=u_{2n}\, \frac{1}{k^2}\, \left(\frac{\Omega(k)}{k^3}\right)^{n-1}\,,\label{rescalingGaussian}
\end{equation}
where $\rho(p^2)$ is the generalized momenta distribution derived from the Wigner law $\mu_W$ \eqref{Wigner}. 
This rescaling adds a linear term in the flow equation, corresponding usually to the canonical dimension, and which in this case depends on the scale. Except for this change, the derivation of the flow equations is fairly standard (see again \cite{Lahoche1}), and we find \footnote{There in an overall factor $2\pi$ regarding the beta-functions considered in \cite{Lahoche1}, due to the convention used in the reference, which multiplied the flow equations by such a factor implicitly because of the definition of canonical dimension. This does not, however, qualitatively change the results.} for $u_2 \ll k^2$ and $k$ small enough:

\begin{equation}
\dot{\bar{u}}_2\approx -2\bar{u}_2-\frac{\bar{u}_4}{18}\,,
\end{equation}
\begin{equation}
\dot{\bar{u}}_4\approx-\mathrm{dim}_{4}\bar{u}_4- \frac{\bar{\tilde{u}}_6 }{15\pi}-\frac{\bar{u}_6 }{30}+ \frac{\bar{u}_4^2}{6}\,,\label{flowu4}
\end{equation}
\begin{equation}
\dot{\bar{{u}}}_6\approx -\mathrm{dim}_{6}\, {\bar{{u}}}_6+\frac{144\bar{u}_4\bar{u}_6}{5}+\frac{8\bar{u}_4\bar{\tilde{u}}_6}{5\pi} -\frac{5\bar{u}_4^3}{9} \,,\label{flowu6}
\end{equation}

\begin{figure}
\begin{center}
\includegraphics[scale=0.6]{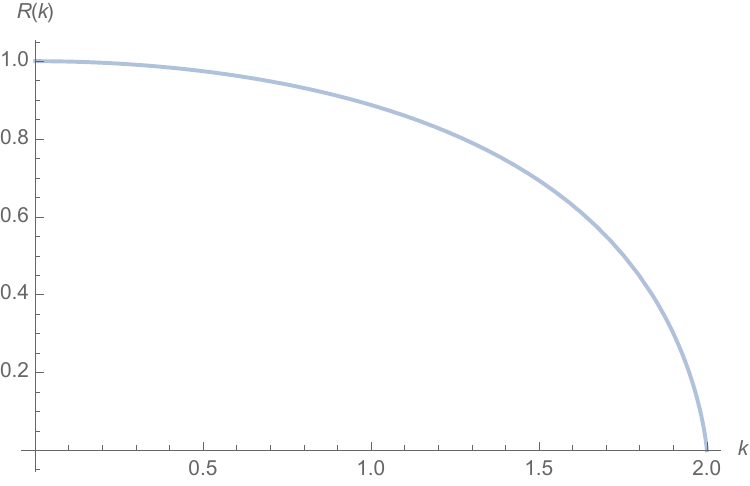}
\end{center}
\caption{Behavior of the factor $R(k)$.}\label{behaviorR}
\end{figure}

Remark that to compute the flow equations before, we neglected some numerical factors whose are numerically of order $1$ for $k$ small enough. For instance, the contribution of order $u_4^2$ in the flow equation for the quartic coupling involves the additional factor:

\begin{equation}
R(k):= k^2 \Omega^{-1}(k) \int dp^2\, \frac{\rho(p^2) \dot{R}_k(p^2)}{(p^2+R_k(p^2))^{5/2}}\,,
\end{equation}

whose behavior is shown on Figure \ref{behaviorR}. For the typical scale where singularities discussed in this paper occur, $k\approx 0.36$ and $k\approx 0.16$, we find respectively $R(k) \approx 0.986$ and $R(k) \approx 0.997$. Numerically, omitting this factor only changes only qualitatively our conclusions, and we took them into account for the numerical investigations of this paper. 

The derivation of the flow equation for the non-local sextic coupling follows the same strategy. The dimension of the coupling is deduced from its contributions in the local sector, and we get (see again reference \cite{Lahoche1}):
\begin{equation}
\dot{\bar{\tilde{u}}}_6=-\mathrm{dim}_{6,n.l}(k)\bar{\tilde{u}}_6\,.\label{behaviortilde}
\end{equation}
where for $\sigma^2=1$:
\begin{equation}
\boxed{\mathrm{\dim}_{2n}(k):= (n-1)\frac{k^2}{k^2-4}+2(2-n) \,,}
\end{equation}
and for the non-local coupling:
\begin{equation}
\mathrm{dim}_{6,n.l}=\frac{k \left(k^2-8\right) \sqrt{4-k^2}+4 \left(8-3 k^2\right) \sin ^{-1}\left(\frac{k}{2}\right)}{\left(k^2-4\right) \left(k \sqrt{4-k^2}-4 \sin ^{-1}\left(\frac{k}{2}\right)\right)}\,.
\end{equation}

\begin{figure}
\begin{center}
\includegraphics[scale=0.6]{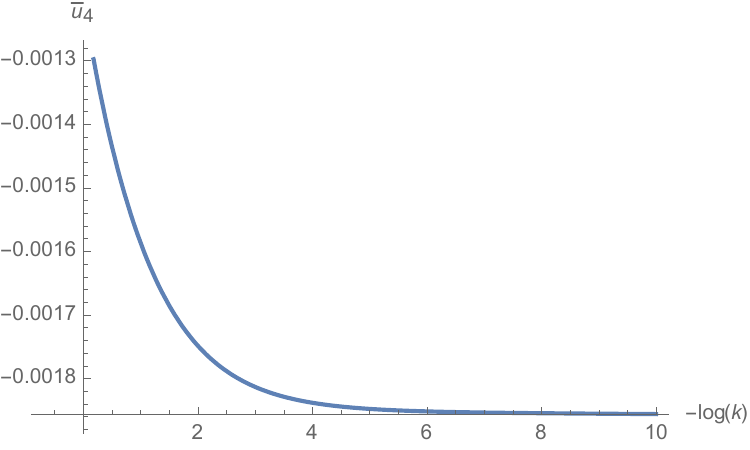}\\
\includegraphics[scale=0.6]{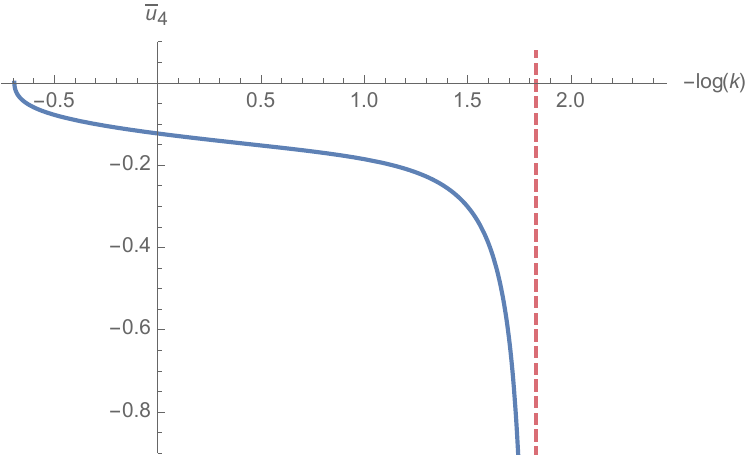}
\end{center}
\caption{On the top: Behavior of the RG trajectories for a small enough value of $\kappa^2$, initial conditions being for $t_0 \approx -0.693$ ($k_0=1.999$). On the bottom, RG trajectories for disorder larger than some critical value $\vert\bar{\tilde{u}}_6(t_0)\vert>\vert \bar{\tilde{u}}_{6,c}(t_0)\vert $. }\label{figbb}
\end{figure}
\begin{figure}
\begin{center}
\includegraphics[scale=0.35]{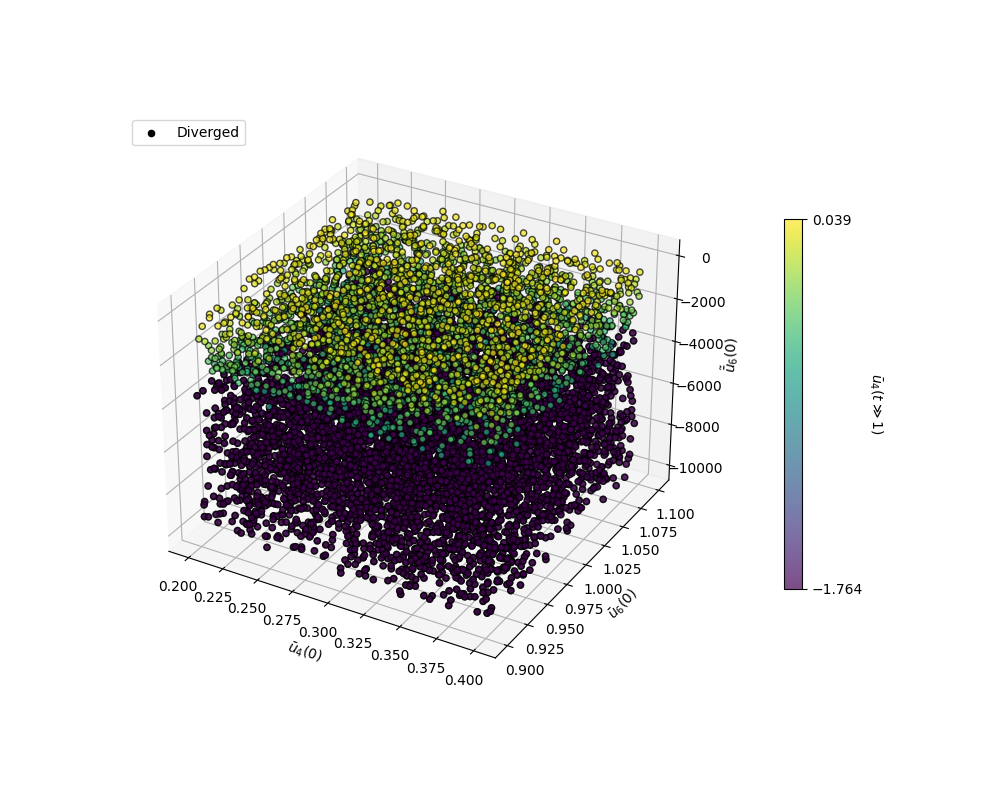}
\end{center}
\caption{Numerical reconstruction of the region of the phase space where finite scale singularities occur. The divergent trajectories are in the black region, which we constructed by setting a fairly low threshold (trajectories that do not diverge generally converge fairly quickly).}\label{fig5}
\end{figure}

Because of the scale dependency of the canonical dimensions and the fact that the non-local sextic interaction does not renormalizes, no global interacting fixed point are expected. Figure \ref{figbb}
illustrates the main point we wanted to emphasize in this section. Sufficiently close to the Gaussian fixed point, when the disorder $\vert\tilde{u}_6 \vert$ is large enough, the trajectories show the appearance of a finite scale singularity reminiscent of the so-called \textit{Larkin
length} \cite{Dupuis,larkin,larkin2}. Moreover, the critical value of the coupling $\tilde{u}_6(t_0)$ as well as the scale at which the singularity appears seems to depend rather weakly on the initial conditions, see \cite{Lahoche1} and Figure \ref{fig5} (the code used to reproduce this figure can be found in \url{https://github.com/IxandraAchitouv/TimeTransInvSymBreakHidden}.). Note that we focused on a particular regime, our objective in this article being essentially to prove a concept.

\section{Landau approximation and phase transition}\label{sec3}

The Landau approximation usually work in the vicinity of the phase transition, assuming that some parameter characterizing the different phase is small enough to expand the thermodynamic functional. We assume that the role of the order parameter is held by the two-point function. This makes it natural to consider the 2-particles irreducible (2PI) formalism and not the traditional 1PI, what is usually the case in the field theory of glassy systems \cite{Dominicis}. The most general 2-points interaction breaking the time translation symmetry reads:
\begin{align}
\nonumber v_{k,1,1}&:=\dfrac{1}{2}\int dt dt^\prime \sum_{\alpha} \gamma(-t,-t^\prime) \textbf{M}_{\alpha}(t)\cdot \textbf{M}_\alpha(t^\prime)\\
&= \frac{1}{2\beta^2} \sum_{\omega,\omega^\prime} \tilde{\gamma}_{\alpha\alpha^\prime}(\omega,\omega^\prime) \tilde{\textbf{{M}}}_{\alpha}(\omega)\cdot \tilde{\textbf{{M}}}_{\alpha^\prime}(\omega^\prime)\,.\label{breakint}
\end{align}
Note that the interaction \eqref{breakint} is not necessary diagonal in the replica space, and we will consider separately to cases:
\begin{align}
\tilde{\gamma}_{\alpha\alpha^\prime}(\omega,\omega^\prime)&= \tilde{\gamma}(\omega,\omega^\prime) \delta_{\alpha\alpha^\prime}\\\nonumber
& \text{or}\\
\tilde{\gamma}_{\alpha\alpha^\prime}(\omega,\omega^\prime)&= \tilde{\gamma}(\omega,\omega^\prime)\delta_{\alpha\alpha^\prime}^\bot\,,\label{case2}
\end{align}

\noindent
where: $\delta_{\alpha\alpha^\prime}^\bot:=1-\delta_{\alpha\alpha^\prime}$ avoids back reaction on the order zero 1PI flow for mass. 

\begin{remark}
We define Fourier transform of some function $f(t)$ as:
\begin{equation}
f(t):= \frac{1}{\beta}\sum_{\omega} \tilde{f}(\omega) e^{-i \omega t}\,,\quad  \tilde{f}(\omega):= \int_{-\beta/2}^{\beta/2} dt \, f(t) e^{i \omega t}\,.
\end{equation}
\end{remark}

\subsection{2-particles irreducible construction}
The understanding of the full momenta dependency of the coupling $\gamma(\omega,\omega^\prime)$ requires more advanced methods than those used in this paper. Because we focus on the vertex expansion at the leading order of the derivative expansion, we can expand the kernel in the power of frequencies:
\begin{equation}
\tilde{\gamma}(\omega,\omega^\prime)=:\Delta \,+\, \delta_1\times(\omega_1+\omega_2)\,+\cdots \,,\label{gammaexp}
\end{equation}
where $\delta_1:=\partial_{\omega_1}\gamma(0,0)$. From the power counting, $\Delta$ has canonical dimension $1$ for all $k$, and $\delta_1$ has dimension zero, higher order derivative contributions have negative dimensions. We can expect that a good enough approximation near the transition can be obtained by considering the most relevant operators. Notice that good enough does not mean ideal, and this is indeed a weakness of this analysis, which we will improve in our future works.

Another source of non-locality for long-time physics could be the following:
\begin{equation}
\tilde{\gamma}(\omega,\omega^\prime)=q \beta (\delta_{\omega 0}+\delta_{\omega^\prime 0})\,.\label{gammaexp2}
\end{equation}
Such an interaction can be generated by some rank one annealed disorder, provided that $q<0$. We will study these two limits in the following. Note that these approximations are by no means exhaustive. For example, we could still consider a term of the form $\propto \beta^2 \delta_{0\omega}\delta_{0\omega^\prime}$, which could be provided by an additional rank $1$ random potential. Such a term however exhibits a somewhat pathological behavior in the continuous limit. It can notably be shown that the back-reaction of the mass term does not cause symmetry breaking as we will observe for the other interactions that we consider here. \medskip 

Within the 2PI formalism, the fundamental quantity is the replicated 2PI effective action $\Gamma_k[\{\textbf{M}_\alpha \},\{G_\alpha\}]$, depending on the 1-points functions $\textbf{M}_\alpha$ and on the 2-points functions $G$ and defined as \cite{Blaizot,Gurau}:
\begin{align}
\Gamma_k[\{\textbf{M}_\alpha \},G]=\frac{1}{2} \mathrm{Tr}\ln G^{-1}+\frac{1}{2} \mathrm{Tr} \, G_0^{-1} G + \Phi[G]\label{2PIfunctional}
\end{align}
where $\mathrm{Tr}$ has the same meaning as in the Wetterich equation \eqref{Wett} and the bare propagator $G_0$ (diagonal in the replica space) is defined as:
\begin{equation}
G_0(\omega^2,p^2):=\omega^2+p^2+m^2+R_k(p^2)\,.
\end{equation}
The last piece in the definition \eqref{2PIfunctional}, $\Phi[G]$ is the so-called Luttinger-Ward functional and expands in term of 2PI diagrams, which in particular determines the gap equation:
\begin{equation}
\Sigma=-2\, \frac{\delta \Phi[G]}{\delta G}\,,\label{gapeq}
\end{equation}
where $\Sigma$ is the standard self energy. The solution of this equation is noting but the so-called Dyson equation:
\begin{equation}
G_k^{-1}=G_0^{-1}-\Sigma\,.
\end{equation}
In the large $N$ limit, the functional $\Phi$ can be computed exactly \cite{Lahoche5,Gurau}, graphically, and for a sextic theory:
\begin{equation}
\Phi[G]\,=\,\vcenter{\hbox{\includegraphics[scale=0.8]{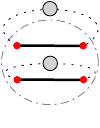}}}\,+\, \vcenter{\hbox{\includegraphics[scale=0.8]{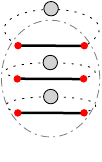}}}\,+\, \vcenter{\hbox{\includegraphics[scale=0.8]{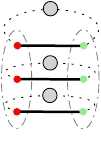}}}\,,
\end{equation}
where the dotted edges with gray discs are propagator $G$, the notation meaning that graphs are computed as Feynman amplitude, replacing the bare propagator with $G$. The 2PI formalism can easily make in contact with the standard 1PI \cite{Blaizot}. Indeed, denoting as $f_k:=\Gamma_k[\{\textbf{M}_\alpha \},G_k]$ the effective action \textit{on shell} i.e. for $G=G_k$, and because $\Gamma_k[\{\textbf{M}_\alpha \}]$ depends  on $k$ only though $G_0$, we get:
\begin{equation}
\dot{f}_k=\frac{1}{2}\mathrm{Tr}\, \dot{R}_k \, G_k\,,
\end{equation}
which is formally equivalent to \eqref{Wett}. This equivalence can be verified for self energy, as well as for higher order 1PI correlations (see \cite{Blaizot} for more details). In the following, we will assume that infinitesimal fluctuations of the 2PI sources can induce correlations breaking the time-translation invariance, and we will assume that the self energy decomposes as:
\begin{align}
\nonumber \Sigma(\omega,\omega^\prime)=\,&\Sigma_N\delta(\omega+\omega^\prime)\\
&-\Delta - \delta_1\times(\omega_1+\omega_2)-q \beta (\delta_{\omega 0}+\delta_{\omega^\prime 0})\,,
\end{align}
where $\Sigma_N$ is the self energy in the \textit{normal phase}, where time translation symmetry works, and satisfies the normal phase gap equation:
\begin{equation}
\Sigma_N\,=\,\vcenter{\hbox{\includegraphics[scale=0.8]{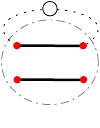}}}\,+\, \vcenter{\hbox{\includegraphics[scale=0.8]{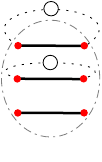}}}\,+\, \vcenter{\hbox{\includegraphics[scale=0.8]{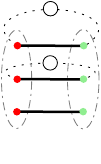}}}
\end{equation}
where blank discs materialize the on-shell effective propagator $G_{k,N}$ in the normal phase
\begin{equation}
G_{k,N}^{-1}=\omega^2+p^2+m^2+R_k(p^2)-\Sigma_N\,.
\end{equation}
This relation is the gap equation for $\Delta=\delta_1=q=0$. As these parameters are non-zero, they induce a back reaction on the effective mass, that we will consider later. Let us notice that in the following, the continuum limit is assumed.

\subsection{Landau expansion}

\paragraph{$q=0$.} To begin, consider the case $q=0$, for the case where $\tilde{\gamma}_{\alpha,\beta}$ is diagonal in the replica space, corresponding to a purely \textit{dynamical ergodicity-breaking} \cite{Baldwin}. From the gap equation \eqref{gapeq}, we find:
\begin{align}
\nonumber\tilde{\gamma}(\omega_1,\omega_2)&\,=\, \vcenter{\hbox{\includegraphics[scale=0.8]{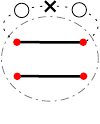}}}\,+\,\vcenter{\hbox{\includegraphics[scale=0.8]{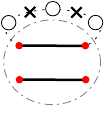}}}\,+\,2\times \vcenter{\hbox{\includegraphics[scale=0.8]{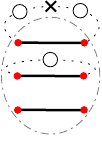}}}\\\nonumber
&+\, \vcenter{\hbox{\includegraphics[scale=0.8]{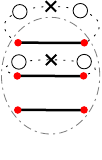}}}\,+\,2\times  \vcenter{\hbox{\includegraphics[scale=0.8]{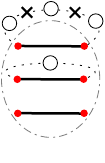}}}\,+\, 2\times  \vcenter{\hbox{\includegraphics[scale=0.8]{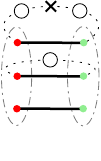}}}\\
&+\, 2\times  \vcenter{\hbox{\includegraphics[scale=0.8]{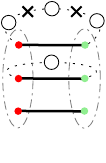}}}\,+\,\vcenter{\hbox{\includegraphics[scale=0.8]{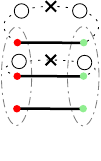}}}\,+\,\mathcal{O}(\gamma^3)\,.
\end{align}
where the cross means a $\gamma$ insertion. The diagrams involved in the expansion \eqref{gammaexp} can be computed using standard perturbation theory, and we get:
\begin{align}
\nonumber \vcenter{\hbox{\includegraphics[scale=0.8]{Phi1bis4O1.pdf}}}&\,=\,-\frac{u_4 \Delta}{3}\,\int dp^2 \rho(p^2) \\
&\times \int d\omega\, G_{k,N}(\omega) G_{k,N}(\omega+\omega_1+\omega_2)\,,
\end{align}
where  $\omega_1$, $\omega_2$ are the external momenta. Note that we omitted the $p^2$ dependency of the propagator to simplify the notations. We will do the same throughout the rest, indicating the integrations but omitting the indices, the structure of the graphs removing any ambiguity. Expanding in power of them, we identify contributions to $\Delta$ and $\delta_1$, the later vanishing because $G_{k,N}$ is expected to be symmetric, we then get:
\begin{align}
\nonumber \vcenter{\hbox{\includegraphics[scale=0.8]{Phi1bis4O1.pdf}}}&\,=\,-\frac{u_4 \Delta}{3}\,\int dp^2 \rho(p^2) (I_{2,0}+\mathcal{O}(\omega_1^2,\omega_2^2,\omega_1 \omega_2))\,,
\end{align}
where:
\begin{align}
&\nonumber I_{m,n}:= \int d\omega\, G_{k,N}^m(\omega) \omega^{2n}\\
&=:C(n,m) (p^2+R_k(p^2)+u_2)^{-n+m-\frac{1}{2}}\,,
\end{align}
where the numerical factor depending on $n,m$ is:
\begin{equation}
C(n,m):=\frac{\Gamma \left(n+\frac{1}{2}\right)  \Gamma \left(m-n-\frac{1}{2}\right)}{ \Gamma (m)}\,.
\end{equation}
Note that we compute the on-shell propagator from the 1PI truncation in the symmetric phase, but the flow for $u_2$ is assumed to be computed including order $\gamma^2$ corrections. For the second diagram, we get in the same way:
\begin{align}
\nonumber \vcenter{\hbox{\includegraphics[scale=0.8]{Phi1bis4O2.pdf}}}\,=\,&\frac{u_4}{3}\,\int dp^2 \rho(p^2) \Big[\big((\Delta^2 I_{2,0}-\delta_1^2I_{2,2}) I_{1,0}\\\nonumber
&-\delta_1^2 I_{1,2}I_{2,0}\big)+\delta_1 \Delta I_{2,0}I_{1,0} (\omega_1+\omega_2)\Big]\\
&+\mathcal{O}(\omega_1^2,\omega_2^2,\omega_1 \omega_2)\,.
\end{align}
In the same way:
\begin{align}
\nonumber \vcenter{\hbox{\includegraphics[scale=0.8]{Phi1bis6O1.pdf}}}&\,=\,-\frac{u_6 \Delta}{15}L_1\,\int dp^2 \rho(p^2) \\
&\times \int d\omega\, G_{k,N}(\omega) G_{k,N}(\omega+\omega_1+\omega_2)\,,
\end{align}
and:
\begin{align}
\nonumber \vcenter{\hbox{\includegraphics[scale=0.8]{Phi1bis6O2_2.pdf}}}\,=\,&\frac{u_6}{15}L_1\,\int dp^2 \rho(p^2) \Big[\big((\Delta^2 I_{2,0}-\delta_1^2I_{2,2}) I_{1,0}\\\nonumber
&-\delta_1^2 I_{1,2}I_{2,0}\big)+\delta_1 \Delta I_{2,0}I_{1,0} (\omega_1+\omega_2)\Big]\\
&+\mathcal{O}(\omega_1^2,\omega_2^2,\omega_1 \omega_2)\,.
\end{align}
where
\begin{equation}
L_n:= \int \rho(p^2) d p^2 \int d\omega\, G_{k,N}^n(\omega)=\int \rho(p^2) d p^2 I_{n,0}\,.
\end{equation}

\begin{align}
\nonumber &\vcenter{\hbox{\includegraphics[scale=0.8]{Phi1bis6O2_1.pdf}}}\,=\, \frac{u_6}{15} \int \rho(p^2)\rho(q^2) dp^2 dq^2 \int d \omega d\omega' d\omega" \\\nonumber
&G_{k,N}(\omega) G_{k,N}(\omega')G_{k,N}(\omega") \Big[ G_{k,N}(\omega+\omega'+\omega") \Big(\Delta^2-\delta_1^2 (\omega+\omega')^2 \\\nonumber 
&+\Delta \delta_1 (\omega_1+\omega_2)\Big)\Big]+\mathcal{O}(\omega_1^2,\omega_2^2,\omega_1 \omega_2)\,.
\end{align}
For simplifying expressions below, we define (we make the dependency over-generalized momenta explicit here for convenience for the reader):
\begin{align}
\nonumber K(p^2,q^2):=\int & d \omega d\omega' d\omega" G_{k,N}(\omega,p^2) G_{k,N}(\omega',p^2)\\
& \times G_{k,N}(\omega",q^2)  G_{k,N}(\omega+\omega'+\omega",q^2)\,.
\end{align}
For the non-local contributions, we get also:
\begin{align}
\nonumber &\vcenter{\hbox{\includegraphics[scale=0.8]{Phi1bis62O1.pdf}}}\,=\, -\frac{\tilde{u}_6}{15} (\Delta+\delta_1 (\omega_1+\omega_2)) \\
&\quad \times \int \rho(p^2) dp^2 \rho(q^2) dq^2 J_{2,1,0}\,+\,\mathcal{O}(\omega_1^2,\omega_2^2,\omega_1 \omega_2)\,,
\end{align}
\begin{align}
\nonumber &\vcenter{\hbox{\includegraphics[scale=0.8]{Phi1bis62O2_1.pdf}}}\,=\, \frac{\tilde{u}_6}{15}\int \rho(p^2) dp^2 \rho(q^2)dq^2  \Big[\Delta^2 I_{1,0}J_{2,1,0} \\\nonumber
&\quad - \delta_1^2 (I_{1,1} J_{2,1,0}+I_{1,0} J_{2,1,1})+\Delta \delta_1 I_{1,0} J_{2,1,0} (\omega_1+\omega_2)\Big]\\
&\quad \,+\,\mathcal{O}(\omega_1^2,\omega_2^2,\omega_1 \omega_2)\,,\label{nonlocalorder2}
\end{align}
where we defined $J_{m,n,p}$ as:
\begin{equation}
J_{m,n,p}:=\int d\omega\, G_{k,N}^m(\omega) G_{k,N}^n(\omega) \omega^{2p}\,,
\end{equation}
the generalized momenta of the $m$ and $n$ propagators are assumed to be different. 
Moreover, in the equation \eqref{nonlocalorder2}, the notations have to be understood as follow: $J_{m,n,p}$ depends on two generalized momenta, and any function being on left (resp. on right) is contracted with the generalized momenta on the $m$ (resp. $n$) propagators. 

For the last diagram, we get finally:
\begin{align}
&\nonumber \vcenter{\hbox{\includegraphics[scale=0.8]{Phi1bis62O2_2.pdf}}}\,=\,\frac{\tilde{u}_6}{15} \Big[\Delta^2 \mathrm{Tr}\, J^2_{1,1,0}-2\delta_1^2 \mathrm{Tr} (J_{1,1,1} J_{1,1,0})\\
&\quad + \Delta \delta_1 (\omega_1+\omega_2) \mathrm{Tr}\, J_{1,1,0}^2\Big]\,+\,\mathcal{O}(\omega_1^2,\omega_2^2,\omega_1 \omega_2)\,,
\end{align}
where here traces means integration over generalized momenta -- see also Appendix \ref{explicit}. Moreover, let us notice that in all these equations, couplings $u_4$, $u_6$ and $\tilde{u}_6$ are assumed to be the dimensioned \textit{bare couplings}. Furthermore, $\tilde{u}_6 \leq 0$. 
\medskip

Computing the integrals involved in these expressions, we find two equations of the form:
\begin{align}
\Delta&=a_1 \Delta+a_2 \Delta^2+ a_3 \delta_1^2+\cdots\\
\delta_1&=b_1 \delta_1+b_2 \Delta \delta_1+\cdots\,.
\end{align}
where the coefficients $a_i$ and $b_i$ depend in particular on the mass $u_2$. At this order, we then get $\delta_1=0$, the state equation for $\Delta$ can be rewritten as an equilibrium condition $U^\prime(\Delta)=0$, with, getting until order $3$:
\begin{equation}
\boxed{U(\Delta):=\frac{1}{2}(1-a_1)\Delta^2-\frac{1}{3}a_2 \Delta^3+\mathcal{O}(\Delta^4)\,.}
\end{equation}
Note that this construction only makes sense because we assume $\gamma$ to be independent (in the sense of functions) of the other terms involved in the construction of $\Sigma$. Another way to discuss the stability of solutions would have been to consider the sign changes of the initial equation of state. It is furthermore easy to check that $\Delta$ has to be positive (for instance, using the formal Dyson formula and discrete Fourier transform). Explicitly we have:
\begin{align}
a_1&:=-\left(\frac{u_4}{3}+\frac{2u_6}{15} L_1 \right)\,L_2\\\nonumber
&-\frac{2\tilde{u}_6}{15} \int \rho(p^2)dp^2 \rho(q^2) dq^2 J_{2,1,0}\\\nonumber 
a_2&:=\left(\frac{u_4}{3} + \frac{2 u_6}{15} L_1\right)\mathrm{Tr}\,I_{2,0}I_{1,0}+\frac{\tilde{u}_6}{15} \mathrm{Tr}\, J^2_{1,1,0}\\\nonumber
&+\frac{2\tilde{u}_6}{15}\int \rho(p^2) dp^2 \rho(q^2)dq^2 I_{1,0}J_{2,1,0}\\
&+\frac{u_6}{15}\int \rho(p^2) dp^2 \rho(q^2)dq^2 K(p^2,q^2)\,\,.
\end{align}


Appendix \ref{explicit} provides some explicit formulas.

\paragraph{$q \neq 0$.} Now, let us move to the different regime, for $\Delta=\delta_1=0$ but $q\neq 0$. In that case, quartic interaction does not contribute, and the expansion has to go at order $q^3$. The relevant diagrams are:

\begin{equation}
\tilde{\gamma}(\omega_1,\omega_2)\,=\, \vcenter{\hbox{\includegraphics[scale=0.8]{Phi1bis62O2_2.pdf}}}\,+\, 2\times \, \vcenter{\hbox{\includegraphics[scale=0.8]{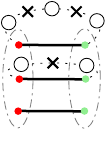}}}\,.\label{closedTrue}
\end{equation}

Computing each diagrams, we get:
\begin{align}
\nonumber \vcenter{\hbox{\includegraphics[scale=0.8]{Phi1bis62O2_2.pdf}}}&\,=\, -\frac{\tilde{u}_6}{15}q^2\, \, \int \rho(p^2) dp^2 \rho(q^2) dq^2 G_{k,N}(0,p^2)\\
& \quad  \times \int d\omega G_{k,N}(\omega,p^2) G_{k,N}(\omega,q^2) G_{k,N}(0,q^2)\,,
\end{align}
and:
\begin{align}
\nonumber &\vcenter{\hbox{\includegraphics[scale=0.8]{Phi1bis62O3_1.pdf}}}\,=\, \frac{\tilde{u}_6}{15}q^3\, \int \rho(p^2) dp^2 \rho(q^2) dq^2 G_{k,N}^2(0,p^2)\\
& \qquad \times \int d\omega G_{k,N}(\omega,p^2) G_{k,N}(\omega,q^2)(p^2,q^2)G_{k,N}(0,q^2)\,.
\end{align}
Finally, the $2PI$ potential looks as:
\begin{equation}
\boxed{V(q):=\frac{1}{2} q^2+\frac{1}{3} c_1 q^3+\frac{1}{4} c_2 q^4+\mathcal{O}(q^5)\,,}
\end{equation}

where:
\begin{align}
\nonumber c_1:=&\frac{\tilde{u}_6}{15}\, \int \rho(p^2) dp^2 \rho(q^2) dq^2 G_{k,N}(0,p^2)\\
&\qquad \times J_{1,1,0}(p^2,q^2)G_{k,N}(0,q^2)\,, \\
\nonumber c_2:=&-\frac{2\tilde{u}_6}{15}\, \int \rho(p^2) dp^2 \rho(q^2) dq^2 G_{k,N}^2(0,p^2)\\
&\qquad \times J_{1,1,0}(p^2,q^2)G_{k,N}(0,q^2)\,.
\end{align}
Note that no back reaction is expected for the 1PI mass flow.

\paragraph{Including replica correlations.} Now, consider the second case, where the time translation symmetry breaking includes correlations between replica (see \eqref{case2}). In that case, the expressions simplify, because only the non-local sextic interaction, the only one coupling different replica, contributes. For approximations \eqref{gammaexp} and \eqref{gammaexp2}, the expansions takes the same form as in \eqref{closedTrue}, and we get respectively:
\begin{equation}
\tilde{U}(\Delta):=\frac{1}{2} \Delta^2+\frac{1}{3}\tilde{a}_1 \Delta^3+\frac{1}{4}\tilde{a}_2 \Delta^4+\mathcal{O}(\Delta^5)
\end{equation}
\begin{equation}
\tilde{V}(q):=:=\frac{1}{2} q^2+\frac{1}{3}\tilde{b}_1 q^3+\frac{1}{4}\tilde{b}_2 q^4+\mathcal{O}(q^5)
\end{equation}
with:
\begin{align}
\tilde{a}_1:=& \frac{\tilde{u}_6}{15} \mathrm{Tr}\, J^2_{1,1,0}\\
\tilde{a}_2:=&-\frac{2 (n-1)\tilde{u}_6}{15} \mathrm{Tr} \,(J_{1,1,0}^2 I_{1,0})\,,
\end{align}
and:
\begin{align}
\nonumber \tilde{b}_1:=&\frac{\tilde{u}_6}{15}\, \int \rho(p^2) dp^2 \rho(q^2) dq^2 G_{k,N}(0,p^2)\\
&\qquad \times J_{1,1,0}(p^2,q^2)G_{k,N}(0,q^2) \\
\nonumber \tilde{b}_2:=&-\frac{2\tilde{u}_6}{15}\, \int \rho(p^2) dp^2 \rho(q^2) dq^2 G_{k,N}^2(0,p^2)\\
&\qquad \times J_{1,1,0}(p^2,q^2)G_{k,N}(0,q^2)\,.
\end{align}

\paragraph{Numerical investigations.} 

As we explained earlier, at order zero in $\gamma$, the RG flow can be computed using the 1PI formalism described in Section \ref{sec2}. Figure \ref{figbb} shows the finite-scale singularity phenomena observed around the Gaussian region for sufficiently strong disorder. For the numerical simulations that follow, we use the same initial conditions as in Figure \ref{figbb}, namely:
\begin{equation} S_0 := {\bar{u}_2(k_0) = 0, \bar{u}_4(k_0) = 0, \bar{u}_6 = 1},, \end{equation}
where $k_0 = 1.999$ is the UV scale. Note that the choice to focus on a specific trajectory does not limit the generality of the results. Furthermore, our goal in this paper is to highlight a specific phenomenon. The initial bar couplings can be computed using the explicit formulas \eqref{rescalingGaussian} and \eqref{behaviortilde}.

\begin{figure}
\begin{center}
\includegraphics[scale=0.55]{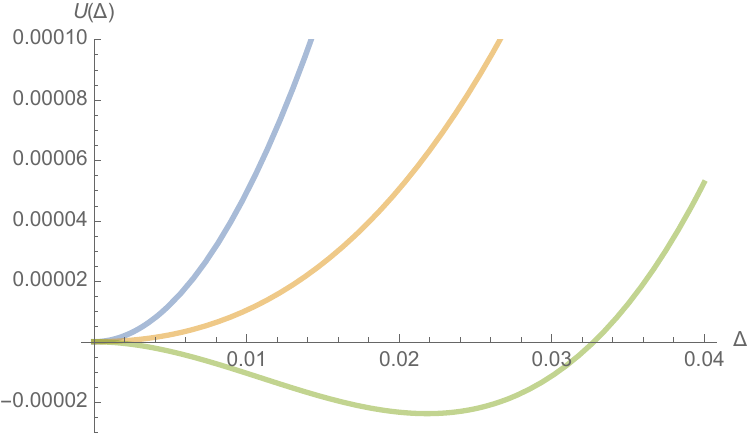}\quad \includegraphics[scale=0.55]{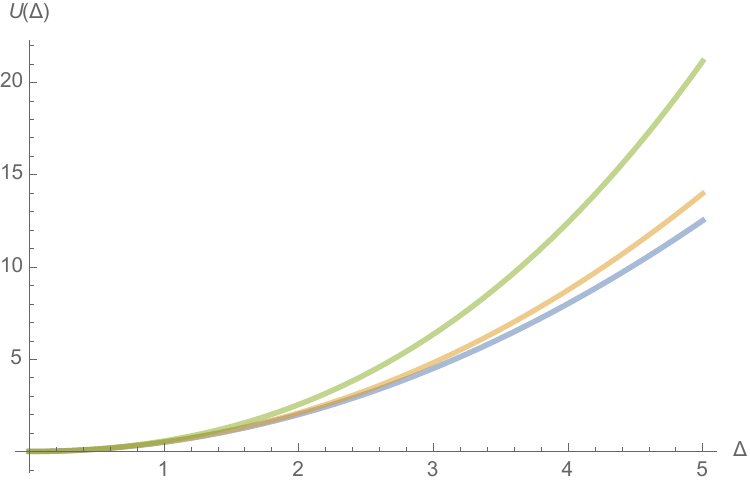}
\end{center}
\caption{On the top: behavior of the potential $U(\Delta)$ for $k=0.36$ (blue curve), $k=0.32$ (yellow curve), $k=0.31$ (green curve). On the bottom: $k=1.64$ (blue curve),  $k=0.36$ (yellow curve), $k=0.13$ (green curve).}\label{figUq}
\end{figure}

\begin{figure}
\begin{center}
\includegraphics[scale=0.55]{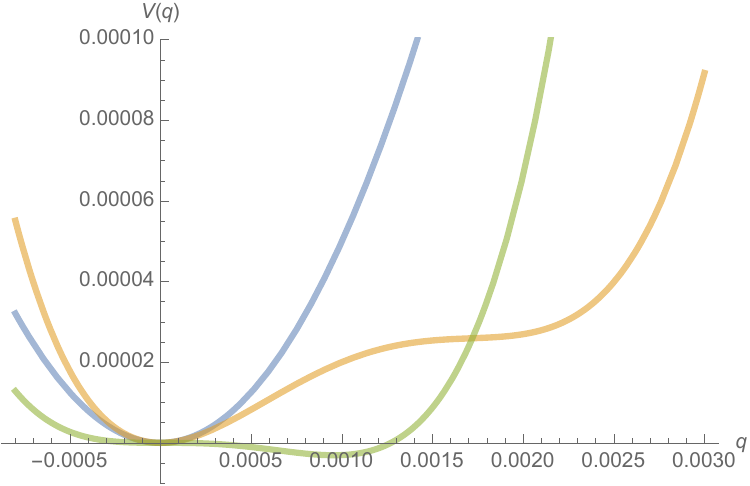}\quad \includegraphics[scale=0.55]{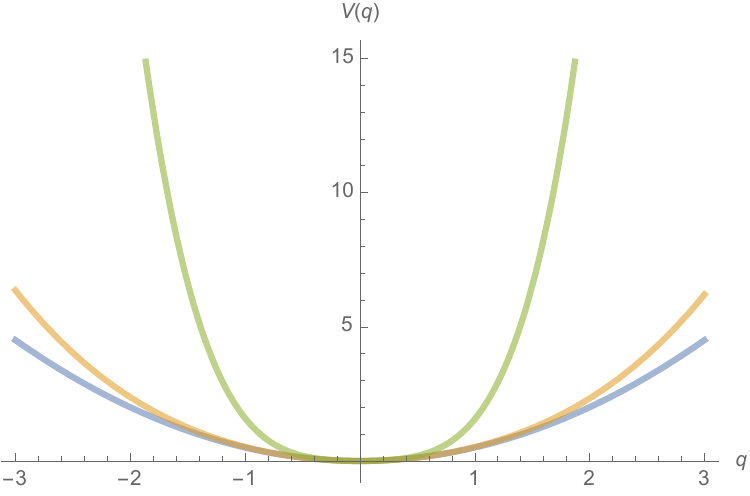}
\end{center}
\caption{Behavior of the potential $V(q)$ for $k=0.36$ (blue curve), $k=0.32$ (yellow curve), $k=0.31$ (green curve). On the bottom: $k=1.64$ (blue curve),  $k=0.36$ (yellow curve), $k=0.13$ (green curve).}\label{figVq}
\end{figure}

\begin{figure}
\begin{center}
\includegraphics[scale=0.55]{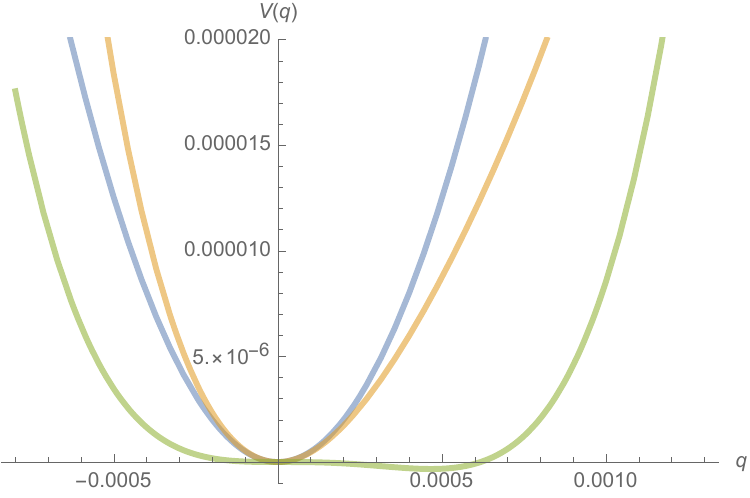}\quad \includegraphics[scale=0.55]{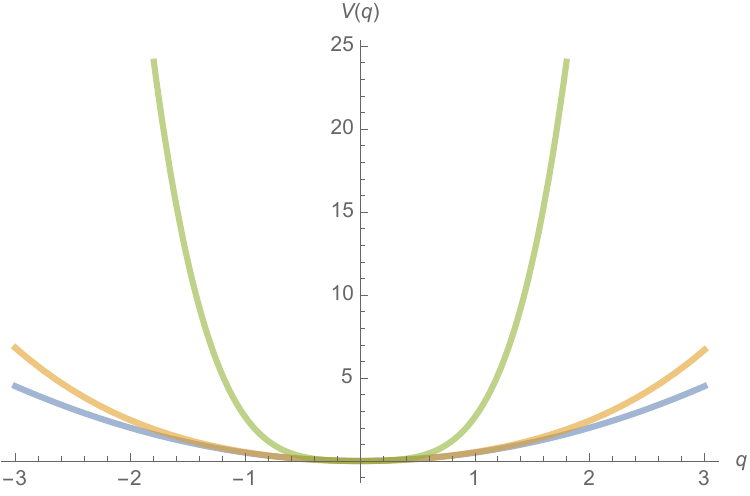}
\end{center}
\caption{Behavior of the potential $\tilde{V}(q)$ for $n=3$, for $k=0.36$ (blue curve), $k=0.32$ (yellow curve), $k=0.31$ (green curve). The second minimum is reaches for $q\approx 0.0005$. On the bottom: $k=1.64$ (blue curve),  $k=0.36$ (yellow curve), $k=0.13$ (green curve).}\label{figtildeVq}
\end{figure}

\begin{figure}
\begin{center}
\includegraphics[scale=0.55]{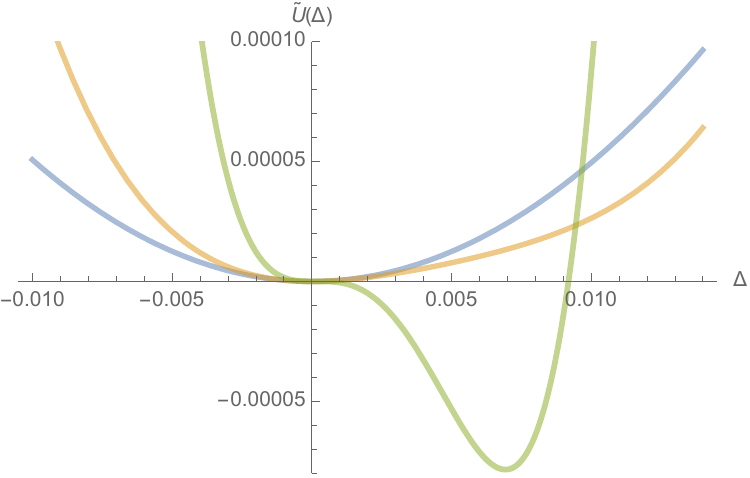}\quad \includegraphics[scale=0.55]{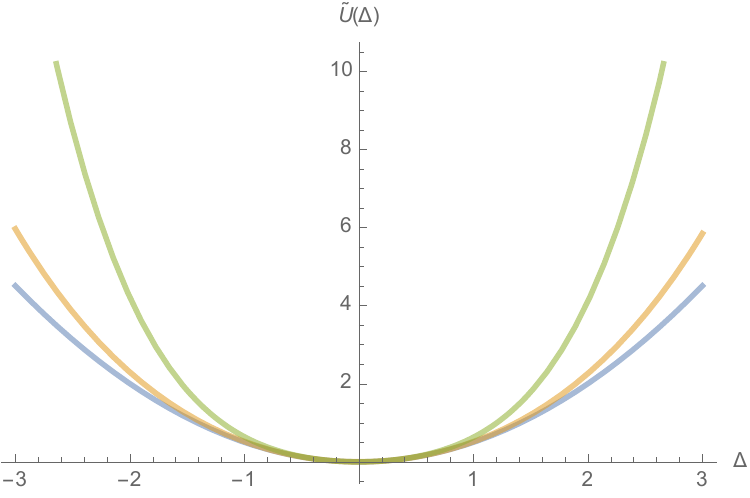}
\end{center}
\caption{Behavior of the potential $\tilde{U}(\Delta)$ for $k=0.36$ (blue curve), $k=0.32$ (yellow curve), $k=0.31$ (green curve). On the bottom: $k=1.64$ (blue curve),  $k=0.36$ (yellow curve), $k=0.13$ (green curve).}\label{figtildeUq}
\end{figure}

\begin{figure}
\includegraphics[scale=0.6]{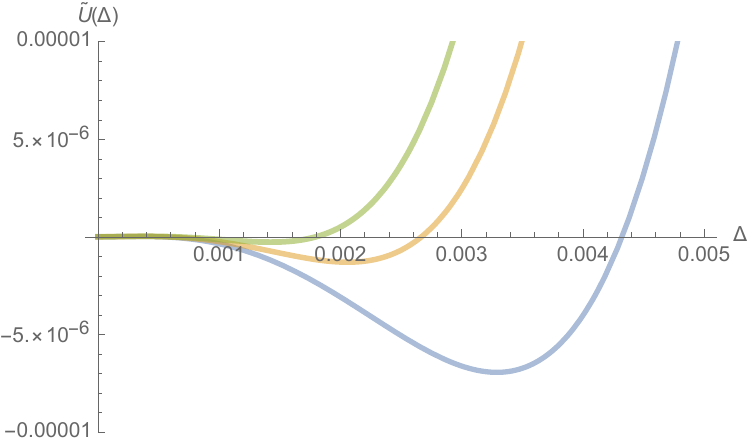}\\
\includegraphics[scale=0.55]{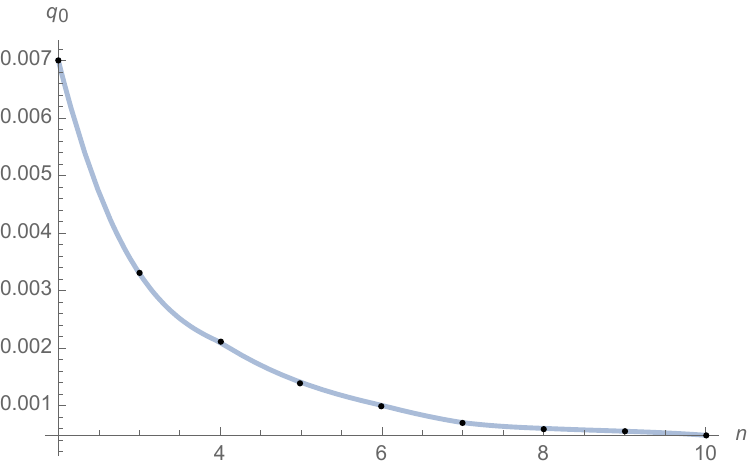}
\caption{On the top: behavior of $\tilde{U}(\Delta)$ for $k=0.31$ and $n=3$ (blue curve), $n=4$ (yellow curve) and $n=5$ (green curve). On the bottom: Dependency of the non-zero minima with $n$. The results seems to suggest a finite limit as $n\to \infty$.}\label{figtildeUq2}
\end{figure}

We consider two values for the initial disorder: $\bar{\tilde{u}}_6(k_0) = -10^3$ and $\bar{\tilde{u}}_6(k_0) = -10$. The first value corresponds to the singular trajectory on the right of Figure \ref{figbb}, while the second corresponds to the singularity-free trajectory on the left. For the bare couplings, these initial conditions correspond to ${\tilde{u}}_6(k_0) = -3.72$ and ${\tilde{u}}_6(k_0) = -0.0372$, respectively. The results are summarized in Figures \ref{figUq}, \ref{figVq}, \ref{figtildeVq}, \ref{figtildeUq} and \ref{figtildeUq2}.

We present the potential for different values of the RG time $t := -\ln k$. The singularity occurs for $t_c \approx 1.83$. The potentials at the top correspond to ${\tilde{u}}_6(k_0) = -3.72$, while those at the bottom correspond to ${\tilde{u}}_6(k_0) = -0.0372$. All the solutions are computed for $n = 2$, except for Figures \ref{figtildeVq} and \ref{figtildeUq2}.
The conclusions are as follows:
1) Phase transitions are observed for all the potentials, except for $V(q)$ due to the back-reaction of the mass flow.
 2)   The phase transition is first order for couplings that are not diagonal in the replica space and second order for diagonal couplings.
  3)  Similar conclusions are observed for higher values of $n$. Increasing $n$ brings the non-zero minimum closer to the origin.
These results should be compared with the analytical insights provided in \cite{Cugliandolo1} and \cite{Cugliandolo2}. 

At this stage, two criticisms can be raised. First, beyond the rudimentary construction method of the RG approximation, we have considered these perturbations separately. It is clear that they are not independent, and we reserve a more comprehensive study of these effects, based on more advanced methods, for future work. Second, one might question the specific form of the couplings we considered. As we have explained, other combinations could have been chosen. However, our goal here was not to find the most optimal form but to provide arguments in favor of a mechanism, which is further supported by analytical insights \cite{Cugliandolo1, Cugliandolo2}.

\subsection{Replica correlations without time symmetry breaking}\label{AppB}

In this section, we summarize the results obtained in \cite{Lahoche1,Lahoche2} concerning the appearance of first-order phase transitions for interactions coupling replicas and corresponding to an equilibrium
ergodicity-breaking. We also illustrate here numerically the fact already conjectured in \cite{Lahoche2} that these metastable states do not appear along all divergent trajectories, and therefore do not systematically allow to cancellation of these divergences.

We consider a self-energy of the form:

\begin{equation}
\Sigma_{\alpha \beta}(\omega,\omega^\prime)=\,\Sigma_N\delta_{\alpha\beta}\delta(\omega+\omega^\prime)-Q \delta_{\alpha\beta}^\bot\delta(\omega+\omega^\prime)\,,
\end{equation}
where $\delta_{\alpha\beta}^\bot:=1-\delta_{\alpha \beta}$ cancel the back reaction of $Q$ one, the zero order 1PI flow for $u_2$. The Landau expansion leads to the potential:
\begin{equation}
U(Q)=\frac{1}{2}Q^2-\frac{1}{3}r_1 Q^3-\frac{1}{4}r_2 Q^4+\mathcal{O}(Q^5)\,,
\end{equation}

where:
\begin{align}
r_1:=&-\frac{\tilde{u}_6}{15}\, \int \rho(p^2) dp^2 \rho(q^2) dq^2\, J_{2,2,0}\,,\\
r_2:=&\frac{2(n-1)\tilde{u}_6}{15}\, \int \rho(p^2) dp^2 \rho(q^2) dq^2\, J_{3,2,0}\,.
\end{align}

\begin{figure}
\begin{center}
\includegraphics[scale=0.55]{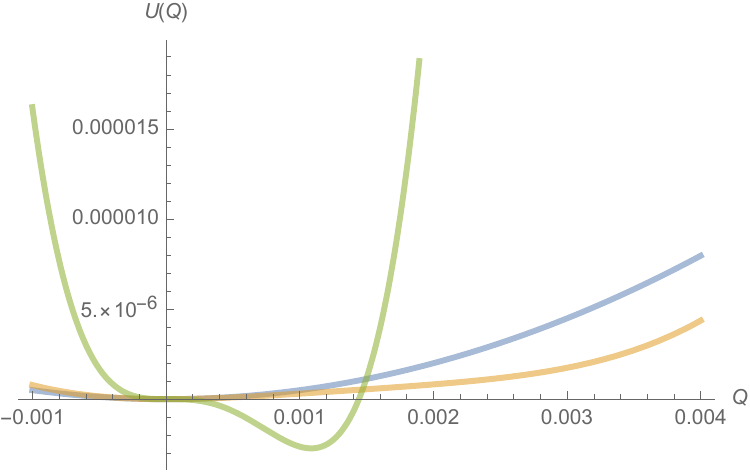}\\
\includegraphics[scale=0.5]{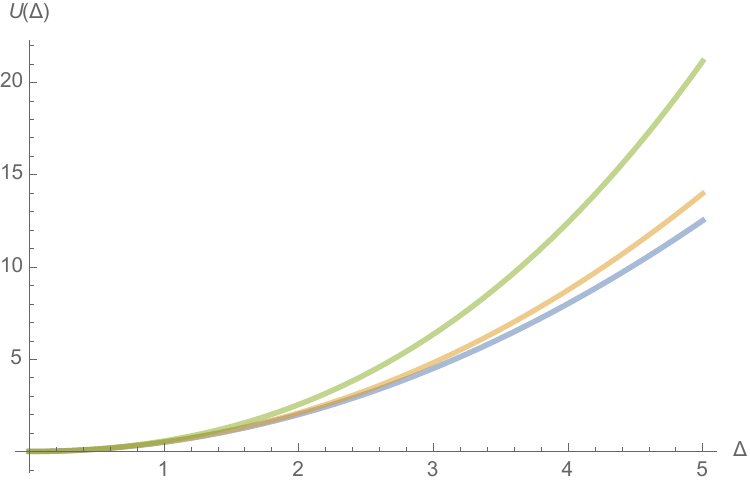}\\
\includegraphics[scale=0.6]{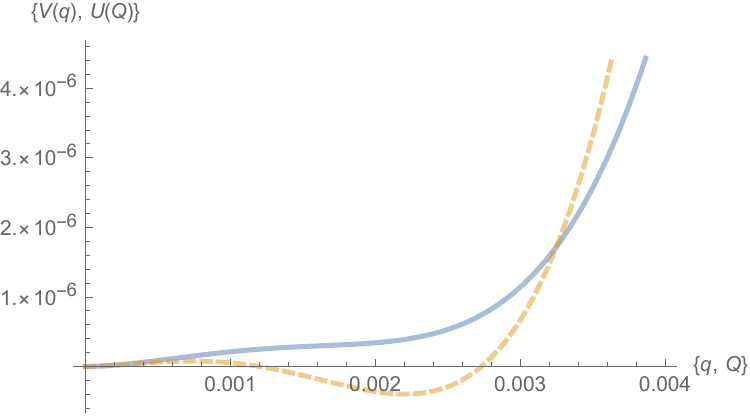}
\end{center}
\caption{Behavior of the potential $U(Q)$ along RG trajectories for $n=2$. On the top for $\tilde{u}_6(k_0)=-3.72$, blue, yellow and green curves are respectively for $k=0.36,0.323$ and $k=0.31$. On the middle for $\tilde{u}_6(k_0)=-0.0372$, blue, yellow and green curves are respectively for $k=1.64,0.36$ and $k=0.13$. On the bottom we show the behavior of potentials $U(Q)$ (blue curve) and $\tilde{V}(q)$ (dashed curve) for $n=2$, at $k\approx 0.321$.}\label{FigAppB}
\end{figure}

Figure \ref{FigAppB} summarizes our conclusions. For the two figures on the top, initial conditions are for $S_0$ and again $k_0 = 1.999$. The two figures at the top and in the middle correspond to the initial condition $S_0$, with $\bar{\tilde{u}}_6(k_0) = -10^3$ and $\bar{\tilde{u}}_6(k_0) = -10$, respectively. We recover the result obtained in our previous work \cite{Lahoche1, Lahoche2}: metastable states appear for some trajectories exhibiting a finite-scale singularity. Taking into account the corresponding interactions, which are forbidden in perturbation theory, cancels this divergence.

The bottom figure compares the behavior of the potential $\tilde{V}(q)$ and $U(Q)$ for som value of $k$, and shows explicitly that phase transition for time translation symmetry breaking appears \textit{before} the one for $Q$. This order however depend on the region of the phase space. In some regions the order can be reversed, or one may happens but not the other. 
\medskip

To conclude, note that similar results are obtained for $n > 2$, with the tendency that the larger $n$ is, closer to the origin is the metastable minima.  

\section{Concluding remarks}

We have assumed that the flow singularities hide a phase transition involving interactions forbidden by perturbation theory. In this paper, we have provided evidence that finite-scale singularities in the ultraviolet flow (i.e., far enough from the mass scale $k^2 f(k) \gg u_2$) can conceal a phase transition that breaks time-translation symmetry. In our previous work \cite{Lahoche2}, we already mentioned that these singularities induce a first-order phase transition involving local correlations between replicas but without breaking any symmetry — see Section \ref{AppB}. Considering these additional interactions appears to resolve these singularities, at least in some regions of the phase space.

The same mechanism seems to occur here: along the singular trajectories, the potential associated with the couplings $\Delta$, derived from the Luttinger-Ward functional, shows the emergence of a stable non-zero vacuum. In this case, however, the transition turns out to be continuous (second order) for diagonal couplings in the replica space. This latter point, seemingly related to the continuous nature of the broken symmetry, strengthens our conclusions by a priori validating a Landau-type expansion of the potential. First-order phase transitions are nevertheless recovered for perturbations that are not diagonal in the replica space.

These findings will be expanded and developed in a forthcoming work based entirely on the 2PI formalism. The goal is to address the current analysis's limitations and reconstruct the system's complete phase space as a benchmark for RG methods in this context.

Finally, it is worth noting that: (1) time-translation symmetry breaking has been considered in other quantum disordered models using different approaches \cite{Time1}; and (2) a typical scenario where time-translation symmetry is broken is the so-called wave function collapse \cite{Time2}, which arises due to the irreversible nature of quantum measurement. As explained in our previous works \cite{Lahoche1, Lahoche2, Lahoche3}, this line of research ultimately aims to tackle the challenging issue of structured disorders \cite{Mezard2}, which is highly relevant for addressing practical quantum information problems.

\appendix

\section{Explicit form of the traces}\label{explicit}

Let us provide some explicit formulas for the traces involved in the expressions of the potentials. For clarity, we include here the explicit dependency on the generalized momenta for the effective $2$-point functions.

\begin{align}
&\nonumber \int \rho(p^2)dp^2 \rho(q^2) dq^2\, J_{2,1,0}\\
&=\int_0^4 \rho(p^2)dp^2 \rho(q^2) dq^2 \int G^2_k(\omega,p^2) G_k(\omega,q^2) d\omega\,,
\end{align}

\begin{equation}
\mathrm{Tr}\, I_{2,0}I_{1,0}=\int_0^4\rho(p^2) dp^2 I_{2,0}(p^2) I_{1,0}(p^2)\,,
\end{equation}

\begin{align}
\nonumber\mathrm{Tr}\, J_{1,1,0}^2&=\int_0^4 \rho(p^2) dp^2 \rho(q^2) dq^2 \int d\omega d \omega^\prime G_k(\omega,p^2) \\
& \times G_k(\omega,q^2) G_k(\omega^\prime,p^2) G_k(\omega^\prime,q^2)\,.
\end{align}

\begin{align}
\nonumber\mathrm{Tr}\, J_{1,1,0}^2 I_{1,0}&=\int_0^4 \rho(p^2) dp^2 \rho(q^2) dq^2 \int d\omega d \omega' d\omega'' G_k(\omega,p^2) \\
& \times G_k(\omega,q^2) G_k(\omega',p^2) G_k(\omega',q^2) G_k(\omega'',q^2)\,.
\end{align}

\begin{align}
\nonumber \int_0^4 \rho(p^2) dp^2 \rho(q^2) dq^2 I_{1,0} J_{210}& = \int \rho(p^2) dp^2 \rho(q^2) dq^2 I_{1,0}(p^2) \\
&\times \int d\omega G_k^2(\omega,p^2) G_k(\omega,q^2)\,.
\end{align}

\begin{align}
\mathrm{Tr}\, J_{2,2,0}= \int_0^4 \rho(p^2) dp^2 \int d\omega \int G_k^2(\omega,p^2) G_k^2(\omega,p^2)\,,
\end{align}

\begin{align}
\nonumber\int \rho(p^2) dp^2 &\rho(q^2) dq^2\,G_k(0) J_{2,2,0}=\int \rho(p^2) dp^2 \rho(q^2) dq^2 \\
&\times G_k(0,p^2) G_k^2(\omega,p^2) G_k^2(\omega,q^2)\,.
\end{align}

\begin{align}
\nonumber \int dp^2 \rho(p^2) I_{20} I_{10}^2&= \int dp^2 \rho(p^2) \int d\omega d\omega'd\omega'' G_k^2(\omega,p^2)\\
&\qquad \times  G_k(\omega',p^2) G_k(\omega'',p^2)\,.
\end{align}

\begin{align}
\nonumber &\int \rho(p^2) \rho(q^2) dp^2 dq^2 K(p^2,q^2) I_{1,0}(p^2) = \int \rho(p^2) \rho(q^2) dp^2 dq^2 \\\nonumber
& \int d\omega d\omega' d\omega'' \int d\omega_1\, G_k(\omega,p^2) G_k(\omega',p^2) G_k(\omega'',q^2)\\
& \times G_{k}(\omega_1,p^2) G_k(\omega+\omega'+\omega'',q^2)\,.
\end{align}

\begin{align}
\nonumber &\int \rho(p^2) \rho(q^2) dp^2 dq^2 I_{10}^2 J_{210}= \int \rho(p^2) \rho(q^2) dp^2 dq^2 \\
&\int d\omega d\omega' d\omega" G_k(\omega,p^2) G_k(\omega',p^2) G_k^2(\omega",p^2) G_k(\omega",q^2)\,.
\end{align}

\begin{align}
\nonumber &\int \rho(p^2) \rho(q^2) dp^2 dq^2 I_{1,0}(p^2) J_{1,1,0}(p^2,q^2)J_{1,1,0}(p^2,q^2)\\\nonumber
&=\int \rho(p^2) \rho(q^2) dp^2 dq^2 \int d\omega d\omega' d\omega" G_k(\omega,p^2) G_k(\omega',p^2) \\
&\qquad \times G_k(\omega',q^2)G_k(\omega",q^2) G_k(\omega",p^2)\,.
\end{align}

\end{document}